1# Transmission System Planning for Integration of Renewable Electricity Generation Units

Amirhossein Sajadi, Luka Strezoski, Richard M. Kolacinski and Kenneth A. Loparo*Abstract*— This paper summarizes the operational challenges imposed by integration of renewable electricity generation units in transmission level where the most common renewable generation units are solar and wind farms at the scale of 100s to 1000s MW. Such units, because of their stochastic nature, introduce new complexity and uncertainty to the grid. Throughout this paper, some results from the recent planning study of integration of 1,000 MW offshore wind farm into the U.S. Eastern Interconnection transmission system are shown.

*Index Terms*— Power System Dynamics, Renewable Energy, Stability, Transmission System## I. Introduction

POWER systems planning is the process of foreseeing load demand, using historic data, and ensuring that sufficient generation capacity expansion and adequate reserve are available throughout 24 hours of planning horizon to meet the demand [1].

Traditionally, the operators had to match the forecasted load and dispatchable generation units. Ever since the advent of power generation by renewable units, new concerns have arisen due to uncertainty and variability of these units. These units could be integrated in both transmission and distribution levels, depending on their size. Wind farms and Photovoltaic farms are usually installed in scales of 100s to 1000s MW which are required to be integrated at the transmission level. As a result of their nature, generation and transmission system planning has become a major concern.

Technical planning of generation and transmission systems is highly correlated. This is because of the fact that the power generated by the generation units is transmitted to the locations where consumers are located by the transmission system. Therefore, they are highly tied and share common stability and control issues.

Technical planning of transmission systems concerns two major areas: (1) Steady state stability and (2) Dynamic stability. The steady state stability of electrical power systems refers to the behavior of system while operating at any given equilibrium operating point. The dynamic stability of electrical power systems refers to the behavior of the system following any disturbance event, by focusing on the trajectory that it takes from pre-disturbance operating point to post-disturbance operating point. Fig. 1 shows the classification of stability studies that are required for assessment of transmission system operation.

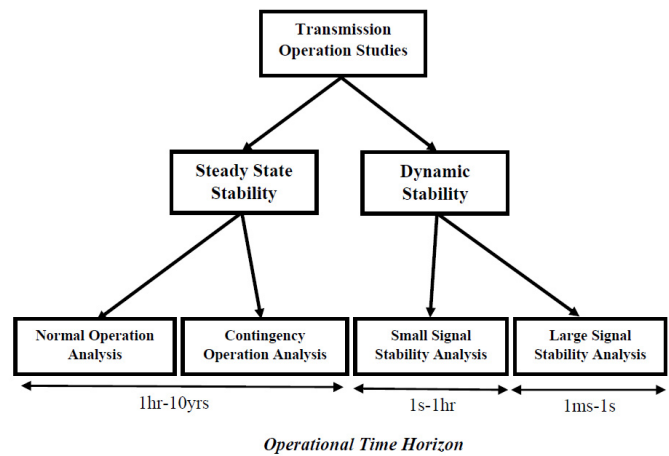

Fig. 1. Classification of transmission system planning studies

This work summarizes the key issues in transmission system planning associated with integration of large scale wind and solar farms. The rest of this paper consists of three main sections: The first section addresses the challenges imposed to the steady state operation of the system. Whereas the second section elaborates the challenges imposed to the dynamics operation of the system. Finally, the conclusion closes this paper. Throughout this paper, some results from the recent planning study of integration of 1,000MW offshore wind farm into the U.S. Eastern Interconnection, FirstEnergy/PJM transmission system, are shown. This study has been carried out by the Case Western Reserve University (CWRU) in collaboration with the team of General Electric (GE), National Renewable Energy Laboratory (NREL), FirstEnergy Utility (FE) and PJM Interconnection (PJM).

## II. Steady State Stability

The steady state operation of transmission systems is associated with two constraints [3]:

1) Line thermal rating of the lines and
2) Voltage regulation across the high voltage system

Usually, the voltage stability margin is much smaller than

This work was supported by US Department of Energy under grant No. DE–EE0005367: Great Lakes Offshore Wind: Utility and Regional Integration Study.

The authors are with the Department of Electrical Engineering and Computer Science, Case Western Reserve University, Cleveland, OH 44106, USA (e-mail: axs1026@case.edu).1

the thermal stability margin. Therefore, the voltage stability margin determines the overall stability margin for the system [4]. Following section discusses the steady state stability related challenges.

*A. Normal Operation*

The fundamental concern of power system planning is to ensure that throughout every 24 hours of everyday of the year, enough generation units are capable and enough line capacity is available to serve the demand load without violation of system's constraint. This is referred to as reliability of the power system. There are three metrics used in industry to determine level of reliability of power systems [5]:

(1) Loss of Load Expectation, LOLE,
(2) Loss of Load Probability, LOLP,
(3) Effective Load Carry Capacity, ELCC.

These are probabilistic indices and their computation requires applying probabilistic methods to the entire interconnected power system model including its constraints such as generation capacities and transmission limits between areas and involve the load profiles, scheduled outages and probabilistic model of forced outages.

The LOLE is expressed in 1 day per 10 years or 0.1 day per year and represents the expected number of the days in a year in which power generation shortage might occur rather than the total outage time.

The LOLP represents, solely, the probability of the event in which power generation shortage occurs without any specific information about the time duration of the shortage. The computation procedure of the LOLE uses the entire daily load profiles when the LOLP uses only the peak values of daily load profiles.

The ELCC is expressed in percentage and represents the contribution of any given generator in capacity of a power system in serving peak load [6], [7].

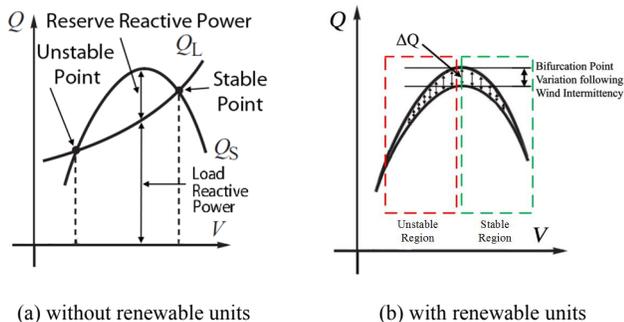

(a) without renewable units     (b) with renewable units
Fig. 1. Steady state voltage stability criteria with and without renewable power generation units.

For integration of a solar or wind farm, due to their highly variable nature, it is required to conduct reliability studies and compute the reliability indices for all levels of contribution of the solar and wind farms in power generation, from 0% to 100% of their capacity. The details of this investigation such as the method used to model the variability of wind or solar farm depends on the operator's preference and experience and the grid architecture. This requires development of more advanced algorithms and analytical tools to quantify the stochasticity of the renewable power generation farms and greater complexity of the power systems.

If the safe reliability level is not reached, additional generation units or reserve units may improve the reliability [1]. Note that after every modification in the system, including change or adjustment in size or status of generation units or load profile, the computations must be repeated until the reliability requirements are met.

*B. Contingency Operation*

Power system security is then defined as the ability of the power system to survive and withstand the contingencies while continuing to deliver power to consumers without interruption [2]. A contingency represents a credible event that can occur in the system such as an outage of a main component or multiple components. Power system security is a broader concept than stability and is assessed by contingency analysis.

Contingency analysis involves applying a sequence of contingency events to the system. These events could be the outage of a transmission line, a generator or multiple components at the same time. The results represent the response of the system to those contingencies. The events are then ranked based on their severity to identify the worst case scenarios and determine how resilient and secure the system will be in the long term. This analysis is usually computationally complex and burdening and may require using multiple processors, parallel computing, and simplified models [4].

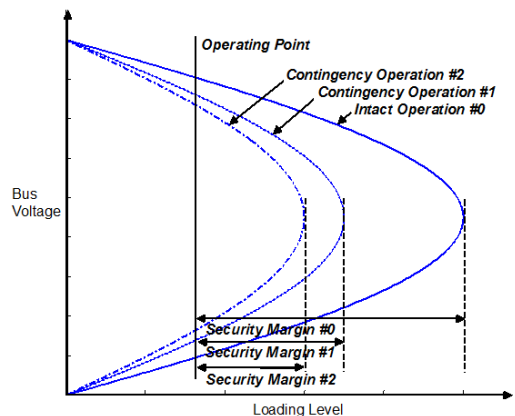

Fig. 2. Visualization of power system security concept

The accuracy of security assessment depends on the severity index used by the planners. The severity index is a metric that quantifies severity of the faults based on how distant the operation point during each fault is from the bifurcation point in which system collapse occurs [9].

The most common approach in power system planning considers *N-1* security criteria in which a system should curtail outage of any single transmission line or generator to stay in the normal or alert operating state [8].

For integration of solar and wind farms, the flow of the power in the system might change. This is because of the fact that these farms inject a significant amount of power into the system and in some cases the non-renewable generators might be required to be redispatched to ensure the balance between generation and consumption. As a result, the security analysis is required to be undertaken for all levels of contribution for the solar and wind farms in power generation, from 0% to 100% of their capacity.

Introduction of stochastic renewable power generation farms escalates complexity and uncertainty in the system therefore, it becomes difficult to find an accurate contingency monitoring algorithm and severity index. The severity indices are meant to identify the pre-contingency and contingency operational states of the system. In the presence of renewable power generation units, both states become uncertain therefore, traditional methods become unhelpful. As a result, new advanced algorithms and procedures are required to incorporate the complexity and size of the system as well as uncertainties of the renewable units.

## III. DYNAMIC STABILITY

The dynamics of a power system concern interactions and behavior of voltage, frequency and rotor angles. The Rotor angle in a synchronous machine is an electrical angle that is defined by relative angle between rotor and stator magnetic fields. In an interconnected power system, power transfers across the transmission lines happen because of the rotor angle difference at the both ends of the line [10]. Therefore, the synchronous generators should operate at a mechanical rotational speed so they produce the same electrical speed. This is known as synchronism in an interconnected power system.

Following a disturbance in the system, rotor speed and, subsequently, rotor angle begin to oscillate. This induces oscillation to active power and reactive power generation. As a result, voltage oscillatory modes and frequency deviation appear in the system. The rotor angle dynamics following a disturbance defined by:

$$\frac{d^2\delta}{dt^2} = \frac{\omega_s}{2HS}\Delta P_{dist} \quad (1)$$

where $\delta$ refers to rotor angle, $\omega_s$ synchronous speed, $\Delta P_{dist}$ disturbance power and $H$ and $S_n$ are inertia constant and MVA rating of the machine. In dynamic studies, it is assumed that prior to a disturbance the system operates at synchronous speed. Therefore, from Equation (1), it can be concluded that dynamic stability of power systems depends on (1) System's inertia and (2) Disturbance power. Inertia is a crucial definition in the dynamics of power systems that describes the mechanical kinetic energy of the generator's rotor at synchronous speed in terms of the number of seconds it takes the generator to provide an equivalent amount of electrical energy, assuming that generate operates at its rated MVA [2].

The oscillations in power system contain several frequencies, known as modes. The crucial modes of the system are electromechanical modes because they can affect tie-line power flows and operational balance in the interconnected power systems [11]. These modes involve generator rotating masses and include Inter-area modes, Interplant modes and Local plant modes and their natural frequency range are 0.1-1.0 Hz, 1.0-2.0 Hz and 2.0-3.0 Hz, respectively. The sources of the electromechanical oscillations usually are shipping bulk power over long distances as well as high gain automatic voltage regulators (AVRs) [12]. The two other modes, Control modes and Torsional modes are referred to as electrical modes. They are excited as a result of mistune of control loops of components and are not in the interest of system operation and planning studies.

Frequency stability in a power system is defined as the ability of the system to adequately manage frequency regulation when disturbances occur [2]. Following a disturbance, when frequency decline occurs, it should be adequately arrested so the under-frequency load shedding (UFLS) relays do not operate.

Primary frequency control, known as inertial response, refers to the response of the system to frequency changes without changing the governors' reference value. The time frame of this stage of frequency control is within the first tens of seconds following a disturbance [2].

Secondary frequency control refers to the action of the balancing reserve units that attempt to mitigate the frequency deviation through Automatic Generation Control (AGC). This control is typically performed by a centralized control system and requires changes in governors' reference value. This stage of control can take up to a few minutes to respond following a disturbance [2].

It is important to note that it takes a few seconds until governor of generators sense the frequency drop and react to it. Therefore, frequency stability of power systems relies on the inertial response to prevent an immediate collapse of the system [21]. The primary frequency response (inertial response) of a multi-machine power system following a disturbance could be expressed by:

$$\Delta f_s = \frac{f_s}{\sum_{n=1}^{i} H_i S_{n_i}} \Delta P_{dist} \quad (2)$$

where $f_s$ refers to synchronous frequency of the system, $\Delta P_{dist}$ disturbance power and $H$ and $S_n$ are inertia constant and MVA rating of every machine connected to the system. From Equation (2), it can be seen that frequency inertial response is highly dependent on inertia of the system and is generated by rotational mass of its generators. In other words, the inertia of a power system is a measure for how the power system resists sudden changes in system frequency [23].

Following section discusses the dynamic stability related challenges.

### A. Small Signal Analysis

Small signal stability analysis addresses the dynamic behavior of the system in respect to the time following a small disturbance and assesses the capability of the system to dampen the oscillations which are caused by the disturbance [2]. In small signal stability analysis, no contingency occurs and all of the grid components are in operation. The disturbances could be, for example, in the form of an expected or unexpected change in the level of generation or consumption.

In practice, the wind and solar power change with a moderate ramp. Solar power could change drastically with a fast pace. However, upon integration of a solar or wind farm, sudden change of level of power generation by the farm could be seen as a disturbance to the system. The effects of this intermittency is assessed through small signal stability analysis for both drop and rise events. This is because wind gusts on many occasions force a sudden shut down of an entire wind farm or a part of it for safety reasons. The generation from solar farms could suddenly drop significantly, depending on geographical specification of it is region. For instance, recently, it was reported that wind gusts of up to 85mph forced shut down dozens of wind turbines all across the UK [24].

The wind and solar farms also must be studied under full extent of their operational range as pre-disturbance operational points. The size of disturbances could be chosen according to the forecasted wind or solar variability for the project which should be found out by energy forecast studies.

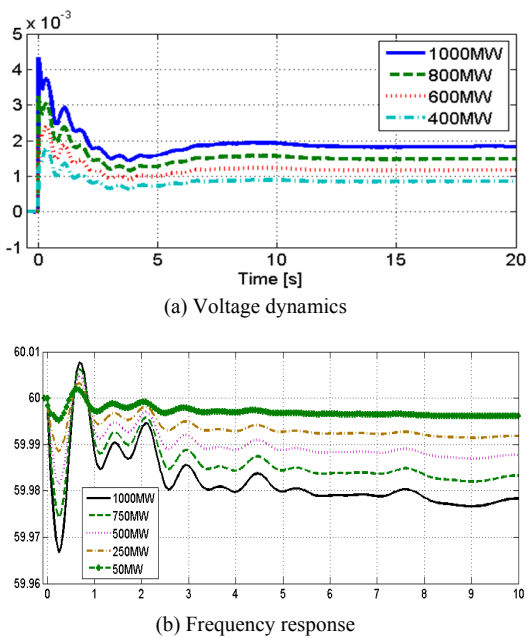

(a) Voltage dynamics

(b) Frequency response

Fig. 3. Voltage and frequency dynamics following the power intermittency

For small signal stability analysis studies, to get a clear and accurate record of oscillation behavior, simulations of 20 seconds or longer are required [13]. This time frame is to ensure all relevant dynamics of the system are captured.

### B. Large Signal Analysis

Large signal stability analysis addresses the dynamic behavior of the system in respect to the time following a large disturbance whether or not it reaches an equilibrium operational point [2]. Following a severe fault, the generators may reset their operating point to a different operating point than prior to the fault. Rotor angle stability in power systems refers to the ability of synchronous machines in the system to maintain their synchronism following a disturbance [2], [25].

The change in rotor angles within a system can change the level and direction of power flow among the lines and generators. This may excite oscillatory modes which should be dampened such that ringing decays within the first few cycles to a few seconds following the fault. Otherwise, the transient behavior of the system may dominate the system response for a sufficient time that the system trajectory diverges from the stability region associated with the current equilibrium point and potentially lead to system-wide failures.

Rotor angle dynamics impact other internal control variables of electrical machines in the system. As a result, voltage and frequency begin to oscillate following a fault. It is necessary to investigate whether they reach their pre-fault or new equilibrium operational points without violating the grid constraints.

The transmission companies have specified technical requirements in their grid codes to assure security and stability of power delivery. The frequency and voltage related requirements are defined by Voltage Ride-Through and Frequency Ride-Through. These requirements are for generation units that ensure they remain in operation during frequency voltage oscillations up to specified time periods and associated voltage levels [26]. This is to prevent unnecessary disconnection of generation units and introduction of additional stress to the grid [27]. These metrics are internationally known definitions and are defined in a majority of grid codes.

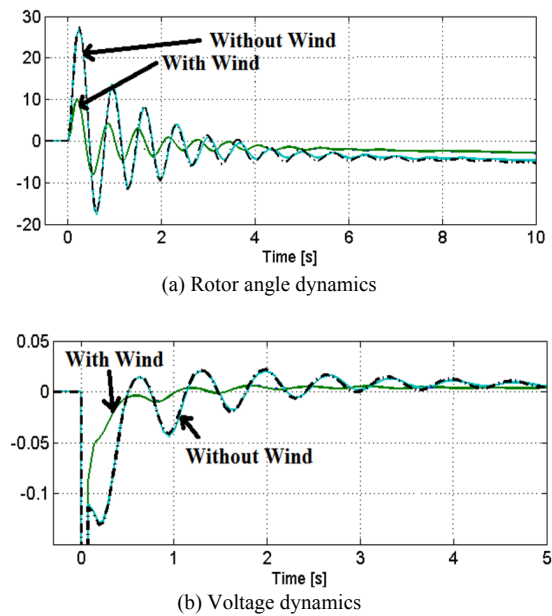

(a) Rotor angle dynamics

(b) Voltage dynamics

Fig. 4. Rotor angle and voltage dynamics following a fault

As was shown in Equation (1), dynamic stability of a power system is highly dependent on the inertia of the system. Wind and solar energy are relatively new sources of power generation in power systems with different dynamic characteristics and behaviors in contrast to conventional generation. Therefore, their integration arises concern in the stability of power systems. This is because generation units such as solar panels and Type 4 wind turbines are connected to the grid through a power electronic interface which makes them fully isolated from the grid while Type 3 wind turbines are partially isolated from the grid [28]. Therefore, their contribution to the grid's inertia is not direct and mechanical. Adversely, it is unreliable, complicated and depends on the controller used in their power electronic interface. In addition, during fault and post-fault reactive power support by them is a function of their characteristic and power electrics capability. This makes them an unreliable addition to a power grid, especially with concern to stability. However, their power electronics interface may use advanced control functions that provide a similar frequency response to the grid as a conventional generator [29].

## IV. Conclusion

This paper summarized the challenges imposed by the renewable power generation units to operation of transmission systems. Steady state stability and dynamic stability related issues upon integration of these units were addressed.

Traditionally, power system operators were meant to match the load with dispatchable generation units. However, in presence of solar and wind farms, due to their uncertainty and variable nature, this becomes more sophisticated and requires my advance algorithms. In addition, these units are often isolated from the grid because of their power electronics interface. This makes their contribution to dynamics of the grid unreliable, complicated and dependent on the controller used in their power electronic interface.